\documentclass{desyproc}

\usepackage{graphicx}
\usepackage{epstopdf}
\usepackage{dcolumn}
\usepackage{verbatim}
\usepackage{float}

\usepackage{hyperref}

\begin{document}
\title{Mississippi State Axion Search: A Light Shining though a Wall ALP Search}

\author{{\slshape Prajwal Mohanmurthy$^1$, Dipangkar Dutta$^{2}$, Joseph Formaggio$^1$, Nicholas Fowler$^{2}$, Mikhail Gaerlan$^{2}$, Yipeng Jiang$^{2}$, John Madsen$^{2}$, Noah Oblath$^1$, Adam Powers$^{2}$, Amy Ray$^{2}$, Robertson Riehle$^{2}$ \\{\bf MASS Collaboration}}\\[1ex]
$^1$Laboratory for Nuclear Science, MIT, 77 Mass. Ave., Cambridge, MA 02139, U.S.A.\\
$^2$Physics Department, Mississippi State University, Mississippi State, MS 39762-5167, U.S.A.}

\contribID{Mohanmurthy\_Prajwal}

\desyproc{DESY-PROC-2014-XX}
\acronym{Patras 2014} 
\doi  

\maketitle

\begin{abstract}
The elegant solutions to the strong CP problem predict the existence of a particle called axion. Thus, the search for axion like particles (ALP) has been an ongoing endeavor. The possibility that these axion like particles couple to photons in presence of magnetic field gives rise to a technique of detecting these particles known as light shining through a wall (LSW). Mississippi State Axion Search (MASS) is an experiment employing the LSW technique in search for axion like particles. The apparatus consists of two radio frequency (RF) cavities, both under the influence of strong magnetic field and separated by a lead wall. While one of the cavities houses a strong RF generator, the other cavity houses the detector systems. The MASS apparatus looks for excesses in RF photons that tunnel through the wall as a signature of candidate axion-like particles. The concept behind the experiment as well as the projected sensitivities are presented here.
\end{abstract}

\vspace{-3mm}
\section{Introduction}
\begin{wrapfigure}{h}{0.5\textwidth}
\vspace{-1.5cm}
\centering
    \includegraphics[scale=.34]{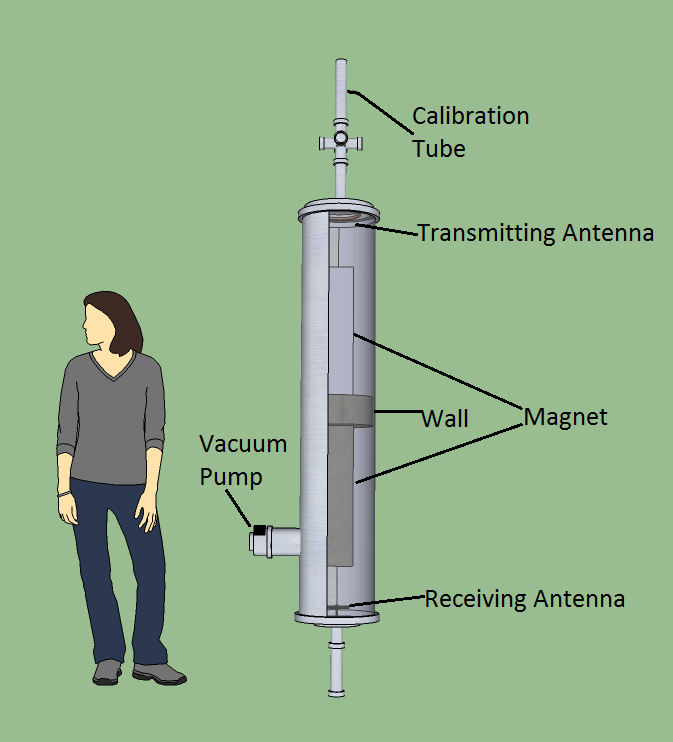}
\caption[]{Basic schematic diagram of the MASS apparatus}
\label{fig1}
\vspace{-7mm}
\end{wrapfigure}

The axion was proposed to solve the strong CP problem \cite{[1]}. Cold axions have been found as a good dark matter candidate \cite{[2]}. In addition to other super-symmetric dark matter candidates, axions are included as dark matter constituents especially in super symmetric extensions of the standard model \cite{[3]}.

A single parameter, the axion decay constant, $f_a$ is sufficient to describe the physics of axions.
\begin{eqnarray}
f_a = 6 \times 10^{-6} eV \frac{10^12 GeV}{m_a}
\end{eqnarray}
where $m_a$ is the mass of the axion. Axions and ALPs are characterized by small masses arising from the shift symmetry of the ALP field, $\phi(x)$. The shift symmetry $\phi(x) \rightarrow \phi(x) + const.$ prohibits explicit mass terms ($\propto m_{\phi}^2\phi^2$) in the ALP Lagrangian. The only way an ALP field could interact with standard model particles is via derivative couplings ($\propto \partial \phi / f_a$). This significantly suppresses their interactions below the $f_a$ scale, effectively making ALPs weekly interacting sub-eV particle (WISP) candidates \cite{[4]}. Furthermore, the two most relevant two photon couplings can be written for both pseudo-scalar and scalar ALPs respectively as;
\begin{eqnarray}
\mathcal{L}_{\phi^{+(-)}\gamma \gamma} = -\frac{g_{+(-)}}{4}F_{\mu \nu} \tilde{F}^{\mu \nu}\phi^{+(-)}
\end{eqnarray}
where, $F_{\mu \nu}$ is the electromagnetic field, $g_{+(-)} = g_{\gamma}(\alpha / \pi f_a)$, $g_{\gamma} \approx -0.97$ in Kim-Shifman-Vainshtein-Zacharov (KSVZ) model \cite{[5]} or $g_{\gamma} \approx -0.36$ in Dine-Fischler-Srednicki-Zhitnitskii (DFSZ) model \cite{[6]} , and $\alpha$ is the fine structure constant.

\vspace{-3mm}
\section{Method}
\begin{wrapfigure}{h}{0.7\textwidth}
\vspace{-1.5cm}
\centering
    \includegraphics[scale=.5]{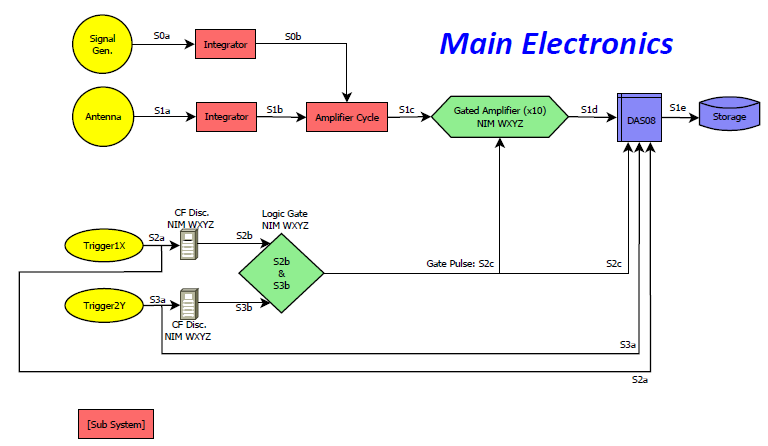}
\caption[]{Schematic diagram showing the fast electronics in line before the data is written to disk.}
\label{fig2}
\vspace{-5mm}
\end{wrapfigure}

In the MASS experiment, Primakoff effect is employed to put the LSW technique to test where the incident light couples with the  magnetic field to create lightly interacting ALPs which pass through an optical barrier, regenerating to photons on the dark side of the barrier, while no photons pass through the barrier. The regenerated photons have the same characteristics as the incident photons, \emph{i.e.} they are of the same frequency, phase and couple to the same electromagnetic mode. Since photons are regenerated via an ALP, the rate of regeneration ($R$) is given by \cite{[7]};
\begin{eqnarray}
R = N_{\gamma}\epsilon_c Q_d P_{\gamma \rightarrow ALP}  P_{ALP \rightarrow \gamma}\\
P_{\gamma \rightarrow ALP} =  P_{ALP \rightarrow \gamma} = \left(\frac{gB}{\frac{m^2}{2\omega}}\right)^2 Sin^2\left(\frac{m^2L}{4\omega}\right)
\end{eqnarray}
where $N_{\gamma}$ is the number rate of photons, $\epsilon_c$ is photon capture efficiency, $Q_d$ is the detector quantum efficiency, $P_{\gamma \rightarrow ALP}$ is the probability of conversion of a photon to a scalar ALP, $P_{ALP \rightarrow \gamma}$ is the probability of conversion of a scalar ALP to a photon, $g$ is the coupling constant, $m$ is the mass of the ALP, $B$ is the magnetic field strength, $\omega$ is the energy of the photons and $L$ is the length of the cavity under magnetic field.

\vspace{-3mm}
\section{Apparatus}
\begin{wrapfigure}{h}{0.55\textwidth}
\vspace{-5mm}
\centering
    \includegraphics[scale=.5]{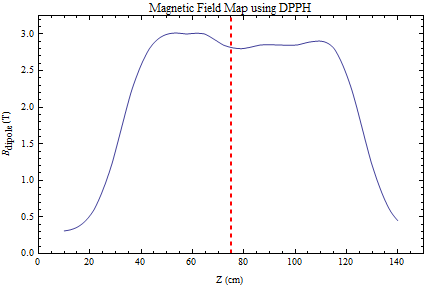}
\caption[]{Plot showing the dipole magnetic field profile in the 2 cavities used in MASS. On the left hand side of the red line is the ``light" cavity and on the right hand side of the red line is the ``dark" cavity.}
\label{fig3}
\vspace{-3mm}
\end{wrapfigure}

The MASS apparatus consists of two tunable evacuated cavities as shown in Figure~\ref{fig1}. These two cavities are mutually optically isolated. The ``light"  cavity houses a StreakHouse $^{TM}$ transmission antenna \cite{[8]}, capable of transmitting integral multiples of 410 MHz and the ``dark" cavity houses an antenna capable of mapping local field in all three spatial axes. A compact dipole magnet bathes the two cavities in a magnetic field in the radial direction while a solenoid is used to create a magnetic field along the axis of the cavities (Z-direction). The radial magnetic field produced by the dipole magnet was mapped using a DPPH (Di-Phenyl Picryl Hydrazyl) probe and is presented in Figure~\ref{fig3}. The solenoid is only used to tune the cavity and thereby has a low field strength output ($<$ 0.5 T). The transmission antenna is mounted at the end of a calibration tube, precise to within 1 nm which allows for additional fine tuning of the cavity. The other end of the calibration tube provides for a number of electronic feed thoughts. Feed through for the receiving antenna electronics is completely separated from the feed through for the main transmission antenna.

The cavities are usually tuned simultaneously to TM$_{010}$ and TM$_{020}$ where the most fundamental mode corresponds to 410 MHz. About 120 W of RF power is dumped into the ``light" cavity through the transmission antenna. The transmission power gives a measure of number rate of photons, $N_{\gamma}$, in Eq. (3). A precise measurement of the standing wave ratio (SWR) of the receiving antenna gives a measure of the product of photon capture efficiency and its quantum efficiency, $\epsilon_c Q_d$, in Eq. (3).

MASS experiment employs a redundant frequency lock-in technique to control signal to noise ratio in the apparatus. The primary signal is generated using a 16-bit programmable digital oscillator which is then amplified a number of times to reach 120 W of continuous transmission power. Both the receiver signal and the primary signal are rectified, integrated and normalized (hereby referred to as processed) such that their difference is null except in the case of there being a regenerated photon (Figure \ref{fig2}). Processing involves continuous rectification which converts all negative components to positive. Furthermore, the rectified signal is integrated over 1 ms before the running integrand is reset to zero. The regenerated photons show up as small excesses in the running integrand at the end of the corresponding 1 ms time period. Integrating the receiver signal also reduces the amount of digitized data by a factor of 410 $\times 10^3$, since the primary signal is 410 MHz and the integrated signal has a frequency of 1 kHz due to the integration time period being 1 ms. The difference between the processed primary signal and the processed receiver signal has the same frequency and phase information as that of the processed primary signal, therefore the processed primary signal also serves as the reference for the three SR 530 \cite{[9]} lock-in amplifiers. The difference of the processed receiver signal and the processed primary signal is then subject to three sets of SR 530 lock-in amplifiers which use the processed primary signal as a reference providing a signal to noise ratio better than $10^{18}$ accounting for residual thermal noise and quantum fluctuations.

The processed and amplified receiver signal is then digitized and written to disk. A set of trigger antenna within the apparatus located around the ``dark" cavity pickup on external noises and creates a gate during which time the processed and amplified primary signal is not written to disk.

\vspace{-3mm}
\section{Conclusion}
\begin{wrapfigure}{h}{0.5\textwidth}
\vspace{-2cm}
\centering
 \includegraphics[scale=.4]{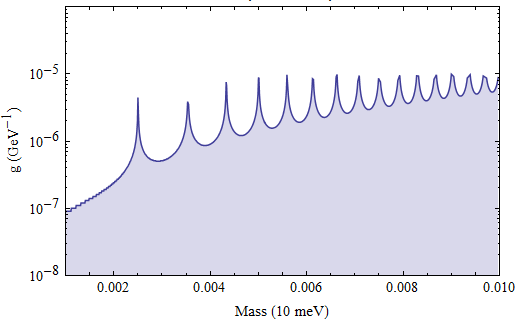}  \includegraphics[scale=.4]{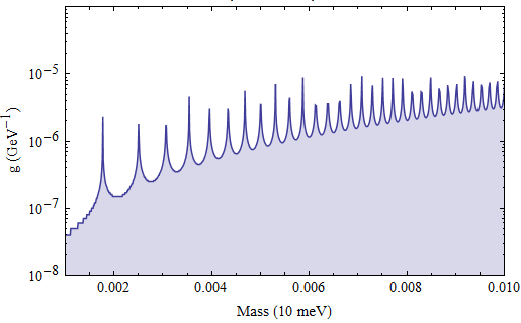}
\caption[]{(Top) Plot showing the sensitivity of MASS to scalar ALPs when the cavities are tuned around 410 MHz with a C.L of 95\%. (Bottom) Plot showing the sensitivity of MASS to pseudo-scalar ALPs when the cavities are tuned around 820 MHz  with a C.L of 95\%.}
\label{fig4}
\vspace{-1.1cm}
\end{wrapfigure}

Taking into account the inhomogeneity of the magnetic field and the characteristics of the two main antennae, one can calculate the possible sensitivity of MASS to scalar and pseudo scalar ALPs as presented in Figure \ref{fig4}. It might be important to note that since the exclusions that MASS can provide for scalar and pseudo scalar ALPs are similar to each other, Figure \ref{fig4} has plotted the sensitivities to two different cavity modes, each a multiple of the fundamental frequency of 410 MHz, both of which are accessible in the MASS apparatus.

Even though the sensitivity that MASS can provide in the low mass regime (10 $\mu$eV - 100 $\mu$eV) for both scalar and pseudo scalar ALPs will only be comparable to the currently available limits, and not any better, it shall demonstrate the feasibility of using the LSW technique using RF photons.

\vspace{-3mm}
\section*{Acknowledgement}
The authors would like to thank Dr. Torsten Clay for his help with setting the group up at the High Performance Computing Collaboratory, the resources of which are used for simulation and analysis. This work is supported by the Mississippi State Consortium and the Jefferson Science Associates through the Thomas Jefferson National Accelerator Facility.


\begin{footnotesize}

\end{footnotesize}

\end{document}